\documentclass[preprint,12pt]{elsarticle}

\usepackage{amssymb}
\usepackage{amsmath}
\usepackage{graphicx}
\usepackage{subcaption}
\usepackage[colorlinks]{hyperref}
\hypersetup{
  citecolor=blue,
  linkcolor=blue,
  urlcolor=blue,
  pdftitle={StiXpert: AI-assisted knowledge discovery and data analysis for solar physics},
  pdfauthor={Hualin Xiao}
}
\usepackage[utf8]{inputenc}
\usepackage[capitalise,noabbrev]{cleveref}

\journal{Astronomy and Computing}

\begin{document}

\begin{frontmatter}

  \title{StiXpert: AI-assisted knowledge discovery and data analysis for solar physics}

  \author[inst1,inst2]{Hualin Xiao\corref{cor1}}
  \ead{hualin.xiao@fhnw.ch}
  \cortext[cor1]{Corresponding author}

  \address[inst1]{University of Applied Sciences and Arts Northwestern
    Switzerland (FHNW), Switzerland}
  \address[inst2]{Physikalisch-Meteorologisches Observatorium Davos/World
    Radiation Center (PMOD/WRC), Davos, Switzerland}

  \begin{abstract}
    Solar-flare studies combine observations and analysis software from multiple
    instruments.
    We present StiXpert, an AI-assisted service developed around STIX for retrieving
    solar-physics knowledge, discovering relevant observations and methods, and
    supporting analysis code.
    Its knowledge base includes solar-physics context, STIX instrument knowledge and
    software, STIX Data Center resources, and selected SSWIDL material spanning multiple
    instruments.
    Its main capabilities are knowledge discovery and assistance with data analysis,
    supporting solar research across instruments.
  \end{abstract}

  \begin{keyword}
    Solar physics \sep Solar flares \sep STIX \sep Artificial intelligence
    \sep Knowledge base \sep Data discovery \sep Scientific software
  \end{keyword}

\end{frontmatter}

\section{Introduction}
\label{sec:intro}

Solar-physics research often combines observations from instruments that differ in energy
range, cadence, observing mode, and data products.
A flare study may bring together hard and soft X-rays, solar energetic
particles, extreme-ultraviolet images, radio measurements, and observations from
spacecraft at different locations.
Researchers must identify which observations cover an event, select suitable products,
and understand the assumptions behind their analysis.

The software used to carry out this work is similarly distributed.
SSWIDL provides a long-standing environment for solar-data analysis
\cite{freeland1998}; the Python ecosystem includes SunPy for solar research
\cite{sunpy2020} and Astropy for general astronomical tasks such as time, coordinates,
and units \cite{astropy2018}.
Instrument teams also maintain dedicated software for accessing, calibrating, and
analyzing their data, using IDL, Python, or other environments.
Moving among these resources can make it difficult to find the appropriate routine,
product, or method, particularly in studies that combine several instruments.

Large language models (LLMs) can answer general questions and assist with code writing,
but their general training does not ensure detailed knowledge of solar phenomena,
instrument capabilities, data products, analysis methods, or the software used to apply
them.
Retrieval-augmented generation (RAG) can ground answers in relevant source material
\cite{lewis2020}, while tool-mediated agents can retrieve and use that information in
successive steps \cite{yao2023}.
Providing AI with this domain-specific knowledge can support scientific knowledge
retrieval, observation and method discovery, as well as code development and analysis
workflows when these are needed.

Here we present StiXpert, an AI service developed around the Spectrometer/Telescope for
Imaging X-rays (STIX), a hard X-ray imaging spectrometer onboard Solar Orbiter that
provides spectral and imaging diagnostics of solar flares \cite{krucker2020}.
StiXpert makes indexed instrument knowledge, data-product information, analysis
software, and related scientific context available to AI clients through retrieval
and tool-based interactions.
Although STIX is the organizing use case, the indexed knowledge also includes selected
SSWIDL resources covering multiple instruments and broader solar-physics
context.
StiXpert is therefore intended to support knowledge discovery as well as code
assistance, and its scope extends beyond STIX.
\section{StiXpert: architecture and interaction model}
\label{sec:architecture}

\subsection{Design objectives}

The system addresses three connected tasks: retrieving relevant scientific knowledge;
using it to identify candidate observations, data products, and analysis methods; and,
when needed, helping the researcher express a selected workflow in the language and
environment they use.
The implementation is anchored in STIX data and expertise, while its indexed SSWIDL
routines, complementary Python packages, and solar-flare and Solar Orbiter context
support workflows that combine STIX with observations from other instruments.
STIX is the organizing use case, not a boundary on the scientific questions and
software that the knowledge base can support.

The design follows four principles.
First, answers to scientific and instrument-related questions should be grounded in
indexed material rather than depend only on model memory.
Second, the assistant should help researchers discover relevant observations, data
products, methods, and supporting software.
Third, any resulting analysis code should be inspectable and run in the researcher's own
environment.
Fourth, the assistant should communicate uncertainty and leave scientific
interpretation and acceptance of results to the researcher.

\subsection{Knowledge and service layers}

\begin{figure}[htbp]
  \centering
  \includegraphics[width=\linewidth]{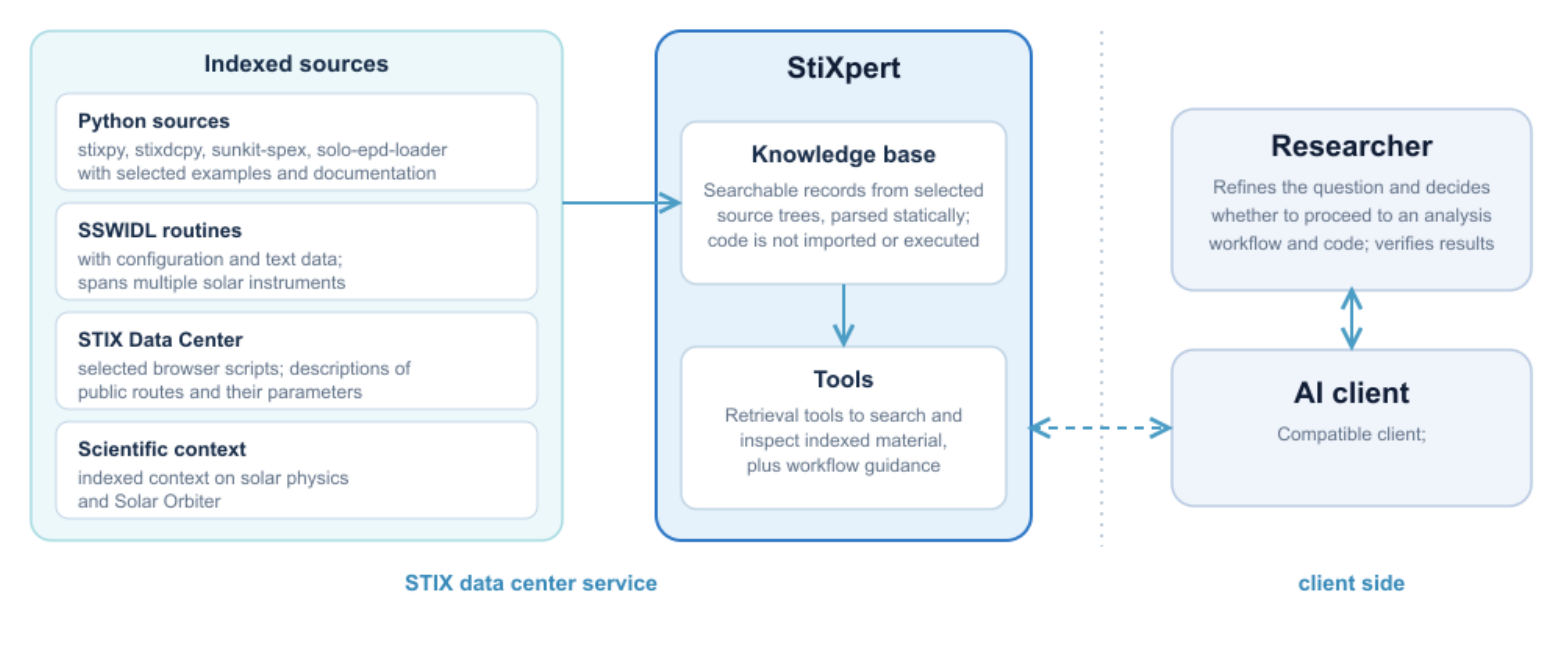}
  \caption{Overview of StiXpert.
    Solar-physics and STIX-specific knowledge, STIX software and Data Center resources,
    and selected SSWIDL resources for multiple instruments are indexed and made
    accessible to an AI client through MCP.
    The client uses these resources to answer knowledge questions, discover observations
    and methods, and support analysis code.
    The server provides retrieval tools; it does not execute analysis or validate
    scientific conclusions.
  }
  \label{fig:architecture}
\end{figure}

\Cref{fig:architecture} summarizes the architecture.
The index builder statically parses selected source trees and stores searchable records
in a knowledge base; indexed code is not imported or executed during indexing.
Python sources include \texttt{stixpy}, \texttt{stixdcpy}, \texttt{sunkit-spex}, and
\texttt{solo-epd-loader}, with selected examples and documentation
\cite{stixpy2026,stixdcpy2022}.
The index also includes SSWIDL routines and supporting configuration and text data,
selected JavaScript running on the STIX Data Center, and descriptions of public Data Center
routes and their parameters \cite{xiao2023stixdc}, as well as indexed context on solar
flares and Solar Orbiter.
This is a software- and source-grounded resource: the indexed SSWIDL material spans
multiple solar instruments, whereas the Python sources and Data Center descriptions
provide stronger STIX-specific coverage.
The combination supports questions that connect STIX with complementary observations
without implying uniform documentation coverage for every instrument.

An MCP server presents retrieval tools and workflow guidance to a compatible AI client.
This gives the model a structured means to search and inspect relevant material rather
than answer solely from its pretrained memory.
In a RAG interaction, a question about a flare can lead first to retrieved information
on its scientific context, instrument coverage, available products, and relevant
methods.
The researcher may then use that information to refine the scientific question or
choose whether to proceed to an analysis workflow and code.
Such retrieval can be useful in its own right; a program need not be generated for the
assistant to help clarify what is known or which observations may address a question.
Retrieved material supports an answer but does not, by itself, establish that a proposed
method is appropriate or that a generated program is correct.

\subsection{From a scientific question to retrieval and analysis}
\label{sec:workflow}

The assistant first interprets the user's prompt to identify the scientific question
and the information or analysis requested.
For a knowledge question, it searches the indexed material and returns relevant
scientific context, publications, instrument descriptions, data-product information,
methods, or software guidance.
This retrieval can answer a question directly; code generation is not required.

When the user requests an analysis, the assistant uses keyword and semantic search to
find the resources most relevant to the prompt.
It first seeks an applicable, predefined workflow.
For example, a STIX flare workflow may identify events, retrieve quicklook and pixel
data, perform spectral analysis with \texttt{sunkit-spex}, and reconstruct images with
\texttt{stixpy}.
For IDL, it searches for corresponding SSWIDL procedures and configuration guidance;
the Python and IDL workflows are related but are not assumed to be identical.
If no predefined workflow is suitable, the assistant searches indexed examples and
source code for relevant procedures or functions, then uses those findings to formulate
an analysis sequence and draft code.

Before code is provided, the assistant can check candidate workflows and routines
against retrieved documentation and examples.
For Python, the reference service also supports static syntax and selected-reference
validation.
These checks do not execute an analysis or establish that its scientific results are
correct: runtime testing with appropriate data must be performed in the user's
environment.
The MCP service itself searches its local index and does not fetch STIX data or run
analysis scripts.

StiXpert can also assist with translating code between Python and IDL, including adapting
an existing routine when no direct counterpart is indexed.
Such translation is a proposed implementation, not evidence of equivalent behavior;
the researcher must check the translated method, inputs, assumptions, and outputs.

\section{Illustrative assistant interactions}
\label{sec:demonstrations}

\begin{figure}[htbp]
  \centering
  \begin{subfigure}[t]{0.49\linewidth}
    \centering
    \includegraphics[width=\linewidth]{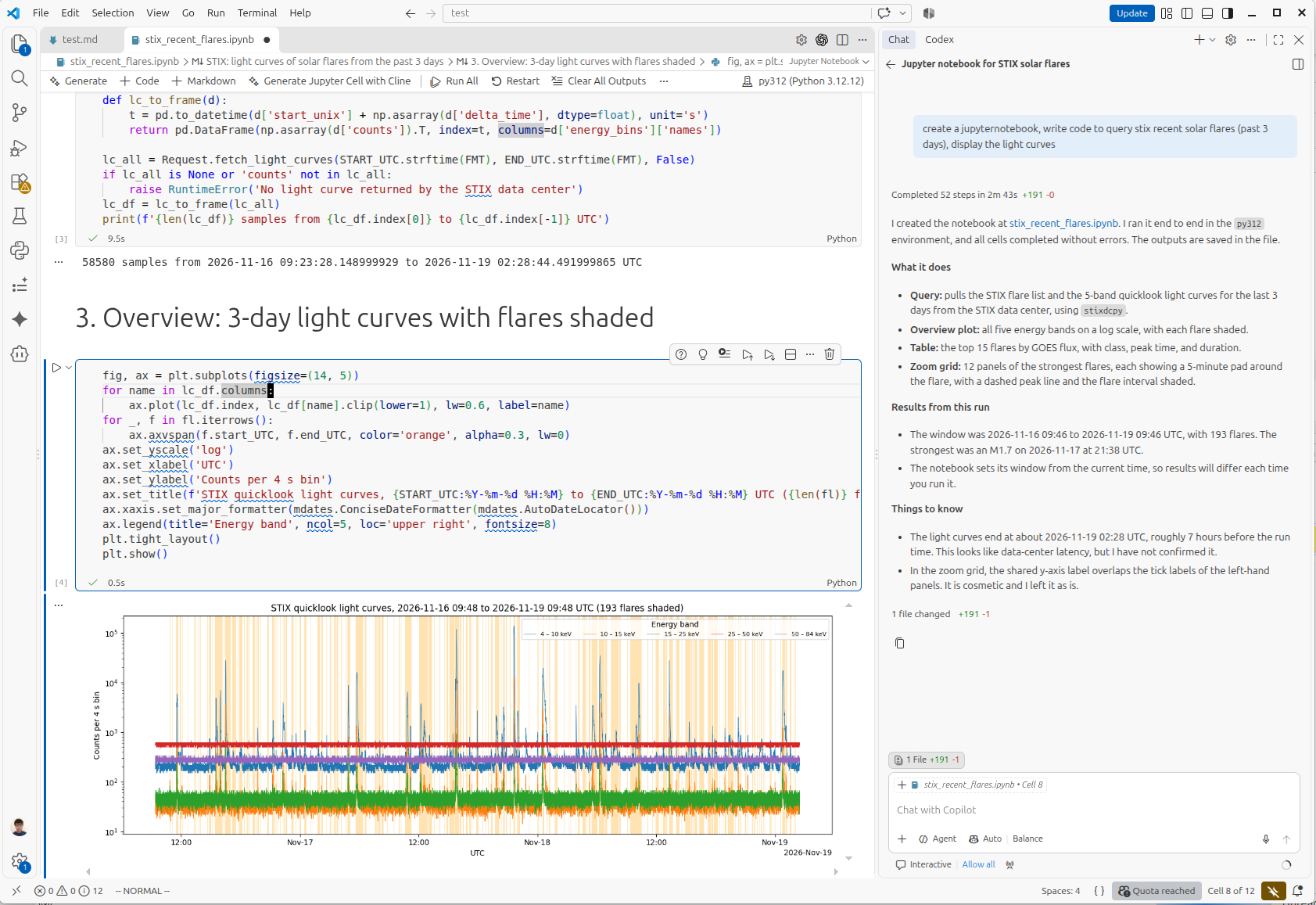}
    \caption{Python notebook example showing assistant-generated code and a STIX
      quicklook light-curve overview.}
    \label{fig:assistant-python}
  \end{subfigure}\hfill
  \begin{subfigure}[t]{0.49\linewidth}
    \centering
    \includegraphics[width=\linewidth]{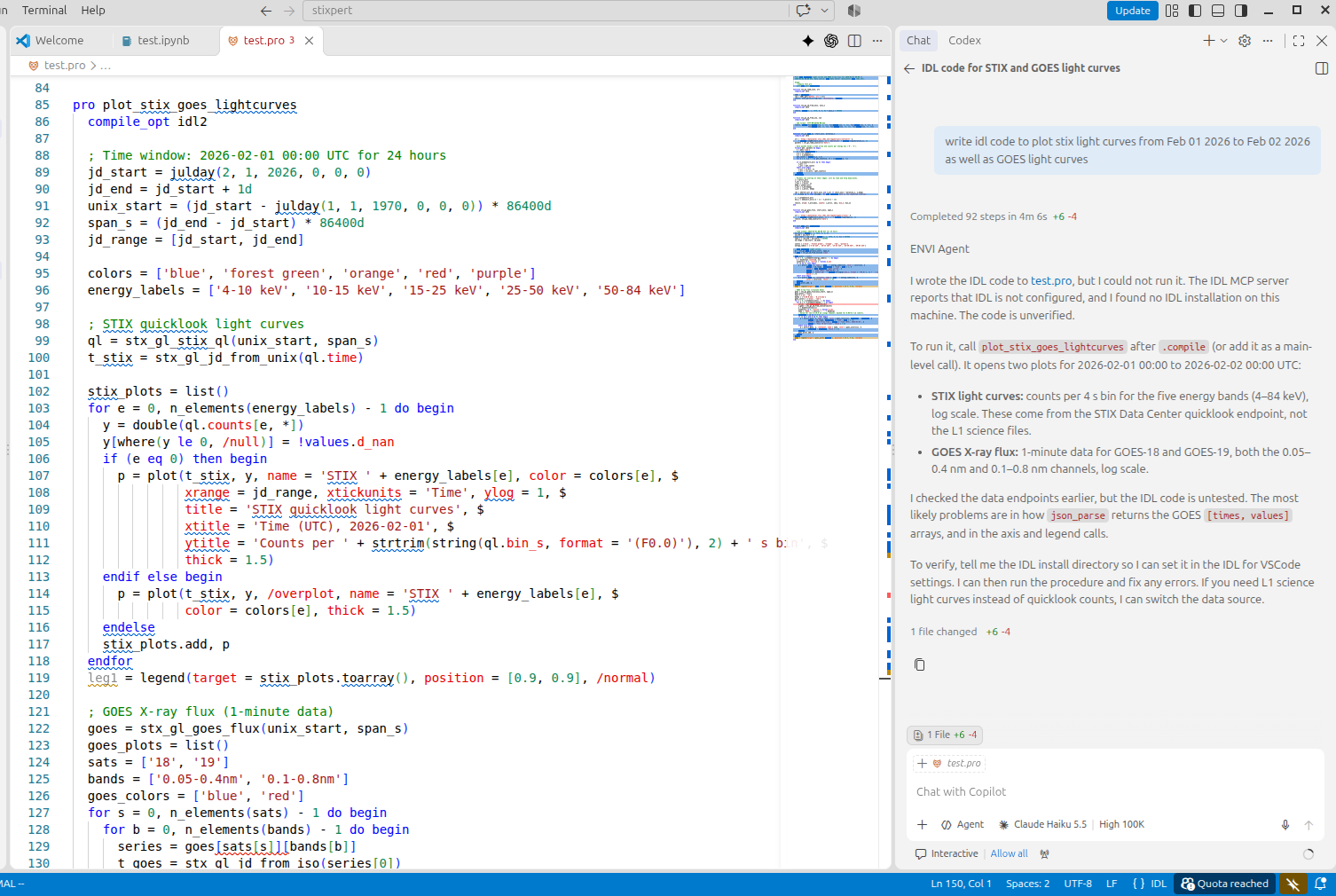}
    \caption{IDL example for plotting STIX and GOES light curves; the assistant
      reports that this code could not be run and remains unverified.}
    \label{fig:assistant-idl}
  \end{subfigure}
  \caption{Examples of solar-analysis code generated with AI assistance in Python and IDL.
    The screenshots illustrate code-generation workflows, not a controlled comparison of
    equivalent implementations: the Python panel shows a plotted result, whereas the IDL
    panel explicitly notes that its code was not executed.
  }
  \label{fig:assistant-code-examples}
\end{figure}

The following examples illustrate code assistance in two analysis environments: a
Python notebook that explores a recent STIX flare list and its quicklook light curves,
and an IDL program draft for a STIX--GOES light-curve comparison.
\Cref{fig:assistant-code-examples} shows the code generated for both prompts.

\section{Discussion and conclusions}
\label{sec:discussion}

StiXpert explores a complementary way to access the knowledge and tools used in solar
research.
Rather than requiring researchers to navigate software, documentation, and data
products as separate resources, a domain-specific AI assistant can retrieve and connect
them in response to a scientific question.
This supports knowledge discovery and can help researchers develop an analysis, with
the aim of reducing software-navigation effort so that more attention can be given to
the physical problem.
The examples demonstrate this interaction pattern, but do not establish improved
research productivity or validate AI-generated analyses.
StiXpert is intended to assist with knowledge discovery and scientific data analysis,
not to replace scientific judgment or authorship.

The approach also suggests a way for instrument teams to provide services to the
research community.
In addition to distributing software, documentation, and data products, teams could
make instrument knowledge and selected tools accessible through an AI-facing MCP
service.
StiXpert is a STIX-organized example of this model, with selected software resources and
solar-physics context supporting questions that involve other instruments.
Instructions for using the service are available at
\url{https://datacenter.stix.i4ds.net/view/research-ai-assistant}.

\subsection{Future perspectives}
\label{sec:future}

Future work should evaluate retrieval quality and analysis assistance on documented
events, including source fidelity, observation coverage, code execution, and agreement
between Python and SSWIDL results.
Extending instrument-facing MCP services to additional observatories could further
support coordinated studies.
In time, AI-assisted discovery may help researchers formulate follow-up questions and
identify promising combinations of observations, while remaining a tool for expert
investigation rather than a substitute for it.

\bibliographystyle{elsarticle-num-names}
\bibliography{bibsample}

\end{document}